\documentstyle[twocolumn,prl,aps,floats,epsfig]{revtex}

\begin{document}
\draft
{\bf Ando, Lavrov, and Segawa Reply:} In the preceding comment
\cite{Comment}, J\'{a}nossy, Simon, and Feh\'{e}r (JSF) offered an
alternative picture to explain our recent findings \cite{ourstr} and argued
that it is unlikely that the charged stripes exist in hole-doped
antiferromagnetic YBa$_2$Cu$_3$O$_{6+x}$ (YBCO). They proposed that the
anomalous behavior of the magnetoresistance (MR) is essentially due to an
in-plane anisotropy in the resistivity of the {\it bulk}. Here we show that
it is extremely unlikely that the charge transport through the bulk is
taking place in the antiferromagnetic YBCO, and thus the formation of
charged stripes in this system is an inevitable conclusion.

The staggered magnetization of the antiferromagnetic (AF) YBCO lies in the
$ab$ plane and there should be two equivalent easy axes for the spins. When
a small number of holes are doped into the CuO$_2$ planes that have the
N\'{e}el order, there are four possible phases for the system: (a) the
system is a single domain AF (all the spins lie along one of the easy axes)
and the doped holes are distributed over the system, (b) an AF domain
structure is formed (with alternating direction of the easy axis) and the
doped holes are distributed in the domains, (c) an AF domain structure is
formed and the doped holes are confined in the domain walls (in other
words, charged stripes constitute the domain walls), and (d) a macroscopic
phase separation takes place into the AF regions and doped non-magnetic
ones.

The ESR experiment by JSF demonstrates \cite{Comment} that there is a
magnetic domain structure, which gives evidence against the case (a). The
equivalence of the $a$ and $b$ axes follows also from the symmetry of the
MR \cite{ourstr}. Case (d) can easily be eliminated, because in this case
the system must contain an admixture of superconducting phase (remember
that YBCO is either an antiferromagnet or a superconductor); experimentally
we have never observed such a macroscopic phase separation. We can also
eliminate case (b), because the domain walls in this case have a very large
energy cost (of the order of $NJ$, where $N$ is the number of
nearest-neighbor spin bonds along the domain walls and $J$ is the
antiferromagnetic interaction energy). Such a large energy cost of the
domain walls would cause a transition into a phase without domain walls
[case (a)] at low temperature. Therefore, the case (c) is the only possible
phase in antiferromagnetic YBCO given the AF domain structure is observed
down to low $T$ \cite{Comment}. In case (c), the energy cost of the domain
walls is grossly smaller than that in case (b) due to the spinless stripes.
The above discussion makes it clear that the charge transport through the
bulk is unlikely to be taking place in the antiferromagnetic YBCO.

JSF argued that an array of charged stripes cannot respond to the magnetic
field because the Coulomb repulsion would make it extremely rigid. This is
their central reasoning to reject the stripe scenario. However, the Coulomb
repulsion does not necessarily make the stripes rigid; actually, the
stripes in cuprates are expected to be fluctuating, meandering, and
liquid-like \cite{theories}.

The question of how the striped structure respond to the external magnetic
field and what is the mechanism for coupling is yet to be clarified. In our
Letter \cite{ourstr}, we suggested some local ferromagnetic moment to be
associated with the stripe structure. To confirm this possibility, we have
recently performed measurements of the bulk magnetization of a
YBa$_2$Cu$_3$O$_{6.3}$ crystal. It turned out that the ferromagnetic
contribution to the magnetization actually exists, since the fits to the
linear high-field $M(H)$ data show noticeable positive intercepts, Fig.~1.
The characteristic field for the ferromagnetic moment to be established,
$\sim$ 4 T, is well correlated with the threshold field in our MR data
\cite{ourstr}. Even a weak hysteresis is observed in the magnetization
curves at 5 K.

To conclude, both the charged stripes and weak ferromagnetism doubted in
the Comment \cite{Comment} should inevitably be taken into account to
understand the nature of antiferromagnetic YBa$_2$Cu$_3$O$_{6+x}$. More
direct investigations would enable us to elaborate on the details of the
striped-phase structure and to understand how exactly the stripes are
affected by the magnetic field.

\begin{figure}[t]
\leftskip-10pt
\epsfxsize=1.05\columnwidth
\centerline{\epsffile{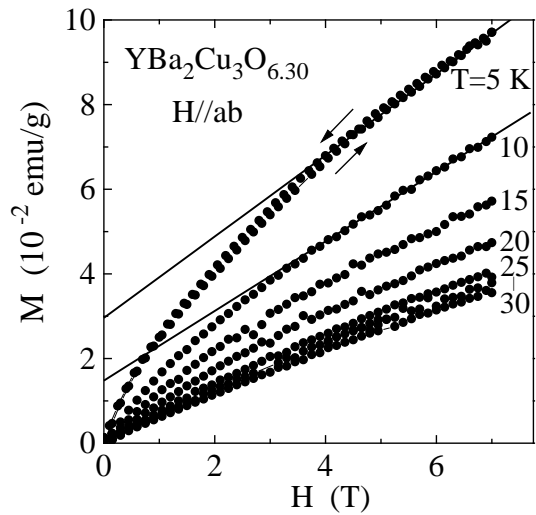}}
\caption{Magnetization of YBa$_2$Cu$_3$O$_{6.3}$ for $H\parallel ab$.}
\label{fig1}
\end{figure}
\vspace{10pt}

Yoichi Ando, A. N. Lavrov, and Kouji Segawa

Central Research Institute of Electric Power Industry,

2-11-1 Iwato-kita, Komae, Tokyo 201-8511, Japan

\medskip
\vfil

\end{document}